\documentclass[aps,prc,twocolumn,showpacs]{revtex4}
\usepackage{graphics}
\usepackage{color}
\usepackage{amssymb}
\usepackage{amsmath}
\usepackage{pifont}
\usepackage{lscape}

\newcommand{\sss}{\scriptscriptstyle}

\begin{document}

\title{$\Sigma$ photoproduction in the resonance region}
\author{S. Janssen}
\email[]{stijn.janssen@rug.ac.be}
\author{J. Ryckebusch}
\author{D. Debruyne}
\author{T. Van Cauteren}
\affiliation{Department of Subatomic and Radiation Physics, \\
Ghent University, Proeftuinstraat 86, B-9000 Gent, Belgium}

\date{\today}

\begin{abstract}
A study of $\gamma p \rightarrow K \Sigma$ processes in an isobar
model at tree level is reported. By comparing model calculations to
the published SAPHIR data, we explore the possible role of different
isospin $I=\frac{1}{2}$ ($N^*$) and $I=\frac{3}{2}$ ($\Delta^*$)
resonances in the reaction dynamics. In our analysis, the inclusion of
the ``missing''  $D_{13} (1895)$ resonance does only slightly improve
the global description of the $\Sigma$ photoproduction data. More
convincing signals for the presence of such a ``missing'' resonance
emerged in the analysis of the isospin related $\gamma p \rightarrow K^+
\Lambda$ reaction. Various implementations of the non-resonant part of
the $\Sigma$ photoproduction amplitude are presented. The sensitivity
of the computed observables and extracted resonance parameters to the
uncertainties inherent to the treatment of the non-resonant
(background) diagrams are discussed.

\end{abstract}

\pacs{13.30.Eg, 13.60.Le, 14.20.Gk, 25.20.Lj}

\maketitle

\section{\label{intro} Introduction}

Traditionally, pion photoproduction reactions $\gamma p \rightarrow
\pi N$ have played a crucial role in studies that aim at elucidating
the excitation spectrum of baryons.  The study of the far weaker kaon
photoproduction channels is currently gaining momentum, thanks to the
construction of a number of dedicated photon and electron accelerator
facilities in the few GeV photon and electron energy regime. It is
believed that the involvement of a $s\overline{s}$ quark anti-quark
pair in the reaction process opens a new window on the dynamics of
excited nucleon states and can help in shedding light on the complex
and not so well understood field of baryon spectroscopy. Recently,
accurate measurements for the three reactions $\gamma p \rightarrow
K^+ \Lambda$, $\gamma p \rightarrow K^+ \Sigma ^0$ and $\gamma p
\rightarrow K^0 \Sigma ^+$ have been reported \cite{Tran,Goers}. In
the baryon resonance region, the observed total cross sections for all
three channels are of the order of a few microbarns.

A property which is specifically related to the production of $\Sigma$
hyperons, is the anticipated role of the isospin $I=\frac{3}{2}$
$\Delta^*$ resonances in the reaction dynamics, which are excluded
from participating in $\Lambda$ photoproduction because of arguments
based on isospin conservation. As such, the $\gamma p \rightarrow K
\Sigma$ channel can act as an isospin filter to study those $\Delta^*$
resonances.  In comparison to the $\gamma p \rightarrow K \Lambda$
process, the description of $\Sigma$ photoproduction within the
context of isobar models appears less attractive, as the freedom to
excite $\Delta^*$ states increases the number of candidates, and
correspondingly the number of parameters, for intermediate resonant
states.  An important feature which helps in minimizing the number of
free parameters, though, is the observation that the $\Sigma^0$ and
$\Sigma^+$ particles are part of the $\Sigma$ isospin
triplet. Consequently, one can rely on isospin symmetry to relate the
coupling constants needed in the description of the $p(\gamma,
K^+)\Sigma^0$ reaction to those required for the $p(\gamma,
K^0)\Sigma^+$ process.  Within such a scheme, a common analysis of
both reaction channels becomes possible.

In this work, we aim at studying $\Sigma$ photoproduction at the
so-called ``tree level'', where only first-order Feynman graphs are
taken into account.  This implies that all higher-order terms, which
for example account for final-state interaction effects, are being
discarded. Also the unitarity constraint is not fully obeyed at tree
level. A recent coupled-channel analysis \cite{Chiang}, specifically
designed to include coupled-channel effects in the description of
strangeness photoproduction channels, reports that the effect of
coupled-channel mechanisms on the cross sections is of the order of
20\%. Admittedly, this is a substantial effect. On the other hand, in
the tree-level analysis presented here, we observe substantial model
dependences. They give rise to variations in some of the extracted
coupling constants up to a factor of two. The major source of the
model uncertainties stems from the treatment of the background or
non-resonant Feynman diagrams.  In the light of this, we reckon that a
profound understanding of the tree-level contributions to hyperon
photoproduction processes is imperative.  We are convinced that
full-blown coupled-channel analyses will also benefit from an improved
understanding of the tree-level contributions to the strangeness
photoproduction reactions.

In this work, we extend our tree-level analysis of the $\Lambda$
photoproduction data reported in Ref.~\cite{Janssen_backgr} to the
$\Sigma$ channel. The organization of this paper is as follows. In
Sec.~\ref{sec:isobar}, we will discuss the isobar model for hyperon
photoproduction on the nucleon. Special attention will be paid to the
peculiar role played by the Born diagrams.  In Sec.~\ref{sec:results}
we then present the results of our numerical calculations. In
\ref{sec:res_contr}, we explore the dominant resonance contributions
and in \ref{sec:back_cont} we investigate various schemes to implement
the non-resonant background terms. In \ref{sec:radcap}, we discuss in
how far the crossing symmetric $K^-p \rightarrow \gamma \Sigma^0$
process provides additional constraints for the description of the
$\Sigma$ photoproduction reaction. Sec.~\ref{sec:concl} contains our
conclusions and an outlook. The Appendix summarizes the use of isospin
constraints in relating hadronic and electromagnetic coupling
constants.

\section{\label{sec:isobar}Isobar Model}

\subsection{\label{sec:form}Formalism}

In this work, the $\Sigma$ photoproduction reaction on the proton is
modeled with the aid of effective Lagrangians. In such a framework,
the physical degrees of freedom are hadrons and their excited states.
Every resonance is treated as an individual particle with its own
properties, like mass, strong-decay widths and photohelicity
amplitudes.  When calculating the leading Feynman diagrams,
effective-field theories provide the operatorial structure of the
interaction vertices and the propagators for the intermediate
particles. The detailed forms of the effective Lagrangians for the
various couplings can be found in many works (see for example
Refs.~\cite{Feuster2,Benmerrouche}). The conventions used here are
summarized in Ref.~\cite{Janssen_backgr}.

Since the Lagrangian formalism describes point-like interactions, it
is a common procedure to introduce phenomenological form factors at
the hadronic vertices.  They do account for the finite extension of
the interacting hadrons and the hard (unknown) physics at short
inter-baryon distances.  We wish to stress that such a phenomenological
treatment of the short-distance physics is necessarily model
dependent. A widely used parameterization for the hadronic form
factors is the dipole form~\cite{Pearce}:
\begin{equation}
F_x \left(\Lambda \right) = \frac{\Lambda^4}{\Lambda^4 + \left(x - M_x^2
\right)^2} \qquad (x \equiv s,t,u) \;,
\label{eq:formfac}
\end{equation}
where $x$ is the off-shell momentum squared at the vertex and
$\Lambda$ is the cutoff mass that sets the short-distance scale of the
effective theory.  It is well known that introducing hadronic form
factors violates gauge invariance at the level of the Born
diagrams. Additional contact terms are then required to restore this
fundamental symmetry, a procedure which is not free of ambiguities.
In our calculations, the gauge-restoring contact terms are determined
with the aid of a procedure which was recently suggested by Davidson
and Workman \cite{Davidson}. The advantage of this recipe over other
schemes, is that the added contact terms do not contain any
singularities.  This is not the case for older procedures, like those
suggested by Ohta \cite{Ohta} or Haberzettl \cite{Haberzettl_gauge},
which have been adopted in numerous theoretical works dealing with
meson induced and meson production reactions.

Within the context of an effective-field theory, the degree of
participation of the different intermediate resonances is determined
by the magnitude of the corresponding electromagnetic and strong
coupling constants. Those values are not predicted by the theory
itself, but should be determined by comparing model calculations to an
extensive data set. The most recently published $p(\gamma,
K^+)\Sigma^0$ and $p(\gamma, K^0)\Sigma^+$ data are due to the SAPHIR
Collaboration at the ELSA facility in Bonn \cite{Tran,Goers}. This
data set for $\Sigma^0$ ($\Sigma^+$) photoproduction contains 21 (5)
total and 70 (18) differential cross section points, as well as 12 (4)
recoil polarization asymmetries over an energy range from threshold up
to 2.0 (1.55) GeV. On the basis of these data, we have determined the
optimal coupling constants for several sets of intermediate resonances
by minimizing $\chi^2$, which is defined in the standard manner:
\begin{equation}
\chi^2 = \frac{1}{N} \sum_{i=1}^N \frac{ \left[ X_i - Y_i\left( a_1,
\ldots a_n \right) \right]^2 }{ \sigma^2_{X_i}} \;.
\end{equation}
Here, $X_i$ are the measured observables, $ \sigma^2_{X_i} $ their
standard deviations, $Y_i\left( a_1, \ldots a_n \right)$ the
theoretical predictions for the variables $X_i$ and $N$ the number of
data points in the fit. The $a_j$'s denote the free parameters
(coupling constants and hadronic form-factor cutoffs) of the model.

The $\Sigma^-$, $\Sigma^+$ and $\Sigma^0$ baryons form an isospin
triplet. As outlined in the Appendix, one can exploit isospin symmetry
to establish relations between the hadronic and electromagnetic
coupling constants for the different $\Sigma$ photoproduction
channels. The obvious advantage of such a procedure, is that the
$p(\gamma,K^+)\Sigma^0$ and $p(\gamma,K^0)\Sigma^+$ data can be
described with one common set of parameters.  In principle, the
$n(\gamma,K^0)\Sigma^0$ and $n(\gamma,K^+)\Sigma^-$ channels could
also be implemented in this scheme \cite{Mart_sigma_1}.  Data for
those reaction channels are sparse, though. Moreover, the procedure of
extracting ``elementary'' neutron cross sections from measurements on
nuclei, like the deuteron, induces severe model dependences.  To make
matters even worse, connecting proton to neutron electromagnetic
coupling constants demands the knowledge of the rather poorly known
helicity amplitudes for the different nucleon resonances.  For all of
the above arguments, we have excluded from our global analyses the
$\Sigma$ photoproduction channels off the neutron.

Even at tree level, the description of $\gamma p \rightarrow K Y$
processes involves a substantial number of Feynman diagrams.  The
diagrams can be divided into two broad classes, the {\em background}
(or, {\em non-resonant}) and the {\em resonant} terms. The latter
reflect themselves as $s$-channel terms and are esteemed to contain
the most relevant physical information. The extracted coupling
constants for the corresponding resonant states constitute the bridge
between the photoproduction measurements and quark-model predictions
\cite{Capstick,Li,Stassart}.  However, the implementation of the
background contributions in the description of $\gamma p \rightarrow K
Y$ processes is far from being a trivial task.  The background
contains several classes of Feynman graphs.  First the Born terms,
involving an off-shell proton in the $s$-channel, a K meson exchange
in the $t$-channel and hyperon exchange in the $u$-channel. Second,
there are terms involving the exchange of a $K^*$ vector meson in the
$t$-channel and $Y^*$ hyperon resonances in the $u$-channel. Some
models dealing with $\Lambda$ and $\Sigma$ photoproduction have
neglected these resonant $u$-channel terms \cite{Mart2,Mart_sigma_2}.
The exchange of $K^*$ and $Y^*$ particles in the $t$- and $u$-channel
are perceived as background and not as resonant diagrams, as they do
not possess poles in the physical region of the reaction.

\subsection{\label{sec:born}Born Diagrams}

\begin{figure}
\resizebox{0.45\textwidth}{!}{\includegraphics{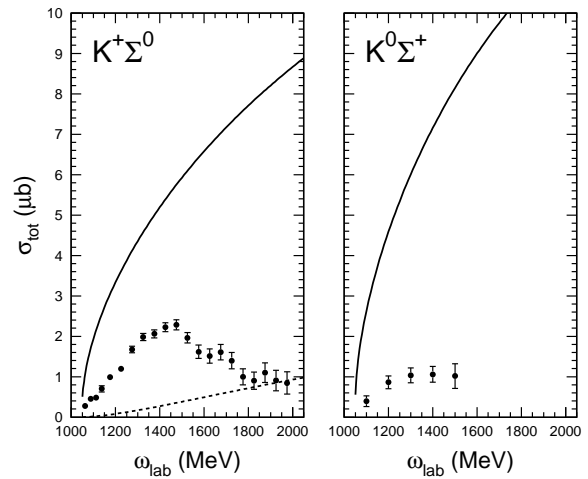}}
\caption{\label{fig:born}The photon-energy dependence of the
contribution from the Born diagrams to the total $p(\gamma,
K^+)\Sigma^0$ and $p(\gamma,K^0)\Sigma^+$ cross sections.  These
results are obtained without introducing hadronic form factors and
$g_{K^+ \Lambda p} / \sqrt{4 \pi}$ = -3.0 and $g_{K^+ \Sigma^0 p}/
\sqrt{4 \pi}$ = 0.9.  The dashed curve in the $p(\gamma,K^+)\Sigma^0$
panel, is the result after excluding the ``extended'' Born term
depicted in Figure \ref{fig:extended}. The data are from
Refs.~\protect\cite{Tran,Goers}.}
\end{figure}

It is a notable fact that the Born terms on their own exhibit rather
intriguing characteristics when it comes to modeling the $\gamma p
\rightarrow K^+ \Lambda$ and $\gamma p \rightarrow K \Sigma$ reaction
dynamics. The Born terms only depend on two parameters, $g_{K^+
\Lambda p}$ and $g_{K^+ \Sigma^0 p}$, which are in principle strictly
related to the well-determined $g_{\pi NN}$ coupling constant through
SU(3) flavor symmetry.  In practice, SU(3) is a broken symmetry and at
best some ``realistic'' ranges for $g_{K^+ \Lambda p}$ and $g_{K^+
\Sigma^0 p}$ can be set. Reasonable ranges, corresponding with
deviations of 20\% from the SU(3) predictions, are:
\begin{subequations}
\label{eq:su3rel}
\begin{eqnarray}
-4.5  \leq  & \frac{g_{K^+ \Lambda p}}{\sqrt{4 \pi}} & \leq -3.0
 \;, \label{eq:su3_l}\\  
0.9   \leq  & \frac{g_{K^+ \Sigma^0 p}}{\sqrt{4 \pi}} & \leq 1.3
 \;.  \label{eq:su3_s} 
\end{eqnarray}
\end{subequations}
Purely hadronic processes, such as $KN$ scattering \cite{Martin} or
$pp \rightarrow p K Y$ reactions \cite{Gasparian,Holzenkamp}, can be
understood in terms of the SU(3)-based coupling constants.  These
observations may suggest that there is little room for SU(3) breaking
beyond the aforementioned ranges.  In the electromagnetic production
of strangeness, though, a difficulty emerges
\cite{Janssen_backgr,Williams,Cheoun,Hsiao}.  Indeed, using coupling
constants which vary within the aforementioned boundaries, the
predicted strength from the ``bare'' Born terms overshoots the
measured $p(\gamma,K^+)\Lambda$ cross section by a factor of three or
four.  In the $\Sigma$ photoproduction case, we observe similar
qualitative features. The photon-energy dependence of the total
$p(\gamma, K^+)\Sigma^0$ and $p(\gamma, K^0)\Sigma^+$ cross sections,
calculated in a model that solely includes the Born terms, is
displayed in Fig.~\ref{fig:born}. For both $\Sigma$ production
channels, the Born terms in themselves strongly overpredict the
measurements.  For the curves of Fig.~\ref{fig:born} we adopt the
values $g_{K^+ \Lambda p} / \sqrt{4 \pi} = -3.0$ and $g_{K^+ \Sigma^0
p} / \sqrt{4 \pi} = 0.9$. They correspond with the smallest absolute
couplings which are allowed according to Eq.~(\ref{eq:su3rel}).  When
adopting couplings based on exact SU(3) symmetry, the overprediction
becomes even more severe.

\begin{figure}
\resizebox{0.30\textwidth}{!}{\includegraphics{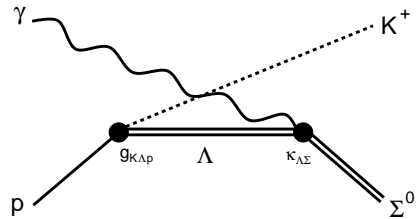}}
\caption{\label{fig:ext_born} The so-called ``extended'' Born term
contributing to the $p(\gamma,K^+)\Sigma^0$ process. A $\Lambda$
hyperon is exchanged in the $u$-channel and converted into a
$\Sigma^0$ at the electromagnetic vertex.}
\label{fig:extended}
\end{figure}

An intriguing observation is that, under the constraints of
Eq.~(\ref{eq:su3rel}), the major fraction of the Born strength in the
$p(\gamma,K^+)\Sigma ^0$ channel is stemming from the so-called
``extended'' Born diagram which is sketched in
Fig.~\ref{fig:ext_born}.
In this diagram, a $\Lambda$ is produced in the $u$-channel and
converted into a $\Sigma^0$ at the electromagnetic vertex. The
dominance of this term can be attributed to the large value of $g_{K^+
\Lambda p}$ compared to $g_{K^+ \Sigma^0 p}$. As a matter of fact,
this implies that the major part of the background cross section in
$\Sigma^0$ production is dominated by $\Lambda$ exchange in the
$u$-channel. Due to charge conservation, $\Lambda$ exchange cannot
contribute to the Born terms for $\Sigma^+$ photoproduction. In the
$p(\gamma, K^0) \Sigma^+$ channel, the Born contributions consist only
of proton exchange in the $s$-channel and $\Sigma^+$ exchange in the
$u$-channel. Remark that also the $t$-channel does not contribute
since the photon does not couple to the neutral $K^0$ meson.  Despite
the restricted number of background diagrams, the two Born terms
produce far more strength than in the $\Sigma^0$ photoproduction
case. This can be partly attributed to the $g_{K^0 \Sigma^+ p}$
coupling constant which is, according to the relation (\ref{eq:SKN}),
a factor of $\sqrt{2}$ larger than $g_{K^+ \Sigma^0 p}$ and partly to
the values of the corresponding anomalous magnetic moments
($\kappa_{\Sigma^+}$ = 1.458, $\kappa_{\Sigma^0}$ = 0.79)
\cite{PDG,Perkins}. 

From the above discussion it may become clear that apart from
introducing resonances, it is mandatory to add model ingredients that
properly counterbalance the strength produced by the ``bare'' Born
terms. In $\Lambda$ photoproduction, where one comes across a similar
difficulty, we discussed three schemes in which this goal could be
accomplished \cite{Janssen_backgr}.  First, it is well-known that for
sufficiently low values of the cutoff mass $\Lambda$ in the hadronic
form factors of Eq.~(\ref{eq:formfac}), the strength stemming from the
Born terms can be adequately suppressed without invoking other
mechanisms.  It turns out that in such a scheme an appropriate
description of the $\Lambda$ photoproduction data can solely be
achieved with cutoff masses of the order of the kaon mass. This is a
rather inconvenient situation for an effective-field theory. Second,
in Ref.~\cite{Janssen_role_hyp} we have shown that for the
$p(\gamma,K^+)\Lambda$ process, the introduction of hyperon resonances
($\Lambda^*$ and $\Sigma^*$) in the $u$-channel permits to naturally
cut down the strength from the Born terms through destructive
interferences.  Such a scheme offers the obvious advantage of
describing the photoproduction data with cutoff masses of a few times
the kaon mass.  As a third alternative, the constraints from (broken)
SU(3)-flavor symmetry of Eq.~(\ref{eq:su3rel}) can be simply
disregarded.  It goes without saying that after sufficiently reducing
the $g_{K^+ \Lambda p}$ and $g_{K^+ \Sigma^0 p}$ couplings, one can
(artificially) force the strength from the Born diagrams to approach
the magnitude of the measured cross sections.

In this work, we extend our description of $\Lambda$ photoproduction
\cite{Janssen_backgr} to the $\Sigma$ production channels. Building on the
knowledge gained in $\Lambda$ photoproduction, in
Sec.~\ref{sec:back_cont} we will present various schemes to implement
the background diagrams in modeling $p(\gamma,K)\Sigma$
reactions. First, however, we will look for a proper set of resonances
to describe the currently available $p(\gamma,K)\Sigma$ data in
Sec.~\ref{sec:res_contr}. Finally, in Sec.~\ref{sec:radcap} we
consider the $p(K^-, \gamma)\Sigma^0$ reaction.

\section{\label{sec:results} Results}

\subsection{\label{sec:res_contr} Resonance contributions}

\begin{table*}
\begin{center}
\begin{tabular}{||c|c|c|c|c|c||c|c||}
\hline \hline
\multicolumn{5}{||c|}{Resonance Contributions} &
Background & $\chi^2$ & {\em NFP} \\ \cline{1-5}
{\em $N^*$core set} & $D_{13}(1895)$ & $S_{31}(1620)$ & $S_{31}(1900)$  &
$P_{31}(1910)$ & Model &\hspace{1.2cm} &  \\
\hline \hline
\ding{70} & & & & & D & 6.52 & 13 \\ \hline
\ding{70} & & \ding{70} & \ding{70} & & D & 4.16 & 15 \\ \hline
\ding{70} & & \ding{70} & & \ding{70} & D & 5.66 & 15 \\ \hline
\ding{70} & & & \ding{70} & \ding{70} & D & 3.20 & 15 \\ \hline
\ding{70} & & \ding{70} & \ding{70} & \ding{70} & D & 3.19 & 16 \\ \hline
\ding{70} & \ding{70} & & & & D & 5.29 & 18 \\ \hline
\ding{70} & \ding{70} & & \ding{70} & \ding{70} & D & 2.88 & 20 \\
\hline \hline 
\ding{70} & & & \ding{70} & \ding{70} & A & 2.03 & 15 \\ \hline
\ding{70} & \ding{70} & & \ding{70} & \ding{70} & A & 1.98 & 20 \\ \hline
\ding{70} & & & \ding{70} & \ding{70} & B & 1.95 & 17 \\ \hline
\ding{70} & \ding{70} & & \ding{70} & \ding{70} & B & 1.81 & 22 \\ \hline
\ding{70} & & & \ding{70} & \ding{70} & C & 1.96 & 15 \\ \hline
\ding{70} & \ding{70} & & \ding{70} & \ding{70} & C & 1.89 & 20  \\ \hline
\hline 
\end{tabular}
\caption{\label{tab:chi}The table summarizes the reduced $\chi^2$
values for the different sets of $N^*$ and $\Delta^*$ resonances and a
particular model to treat the background diagrams. The $\chi^2$ values
are from the best fits obtained from comparing the model calculations
with the SAPHIR $p(\gamma,K)\Sigma$ data. With {\em ``$N^*$core set''}
we refer to the $S_{11}(1650)$, $P_{11}(1710)$ and $P_{13}(1720)$
nucleon resonances. {\em ``NFP''} indicates the total number of free
parameters in the corresponding fitting procedure.}
\end{center}
\end{table*}
Recent isobar models
\cite{Feuster2,Janssen_backgr,Mart2,Saghai_hyp_res} identified the
three nucleon resonances $S_{11}(1650)$, $P_{11}(1710)$ and
$P_{13}(1720)$ as leading $N^*$ contributions to the $p(\gamma,
K^+)\Lambda$ reaction.  It thus appears natural to consider them as
privileged candidates to participate in the $\Sigma$ photoproduction
channels \cite{Mart_sigma_2}. On the other hand, we do not have
similar guidelines concerning the leading $\Delta^*$ contributions. In
our numerical investigations, we rely on a $\chi^2$ procedure to judge
whether a particular set of resonances is suited to describe the data.
In the process of determining an optimum set of $N^*$ and $\Delta^*$
particles, we have fixed a basic set consisting of the three
aforementioned $N^*$ resonances to which we have gradually added other
combinations of $N^*$ and $\Delta^*$ states. All results reported in
this subsection are obtained with a particular model choice (in
Sec.~\ref{sec:back_cont} coined model D) for treating the
background. A profound discussion of the implementation of the
background diagrams is postponed to Sec.~\ref{sec:back_cont}.

Starting with a ``core set'' consisting of the $S_{11}(1650)$,
$P_{11}(1710)$ and $P_{13}(1720)$ resonances, we arrive at $\chi^2$ =
6.52 for an overall fit to the combined set of $p(\gamma,K^+)\Sigma^0$
and $p(\gamma,K^0)\Sigma^+$ cross section and polarization asymmetry
data.  This quality of agreement surely allows room for improvement
and, consequently, for additional $N^*$ and $\Delta^*$ resonances
playing a non-negligible role in the reaction
dynamics. Table~\ref{tab:chi} summarizes the attained $\chi ^2$ values
for various combinations of resonances.  In an attempt to minimize the
number of free parameters, we started out with introducing only
spin-1/2 $\Delta^*$ states. Note that in an effective Lagrangian
approach, a spin-1/2 resonance adds only one free parameter while five
extra parameters are introduced per spin-3/2 resonance (two coupling
constants and three off-shell parameters). Candidates for spin-1/2
$\Delta^*$ resonances are the $S_{31}(1620)$, $S_{31}(1900)$ and
$P_{31}(1910)$ states \cite{PDG}. With those three $\Delta^*$
resonances and the aforementioned core of three $N^*$ resonances
consisting of the $S_{11}(1650)$, $P_{11}(1710)$ and $P_{13}(1720)$,
we arrive at a global best fit with $\chi^2$ = 3.19.  A similar
quality of agreement ($\chi^2$ = 3.20), however, can already be
achieved by the mere action of only two of these $\Delta^*$'s, the
$S_{31}(1900)$ and $P_{31}(1910)$. Other combinations selected out of
the three aforementioned $\Delta^*$ states were also able to improve
the description of the data (see Table~\ref{tab:chi}) although the
combination of the $S_{31}(1900)$ and $P_{31}(1910)$ clearly produced
the best $\chi^2$.  Note that these two $\Delta^*$ resonances were
also recognized as most likely $I = \frac{3}{2}$ resonance candidates
by Mart in his analysis of the $\Sigma$ photoproduction data
\cite{Mart_sigma_2}.

The recent $p(\gamma,K^+)\Lambda$ data from the SAPHIR collaboration
\cite{Tran} exhibit a structure in the energy dependence about
$\omega_{lab} \approx$ 1.5 GeV.  In the analysis of Ref.~\cite{Mart2},
this structure was put forward as possible evidence for the existence
of a ``missing'' $D_{13}(1895)$ resonance.  An alternative
interpretation in terms of the exchange of a spin-3/2 $\Lambda^*$
resonance in the $u$-channel was put forward by Saghai in
Ref.~\cite{Saghai_hyp_res}.  In Ref.~\cite{Janssen_backgr}, we
stressed that $N^*$ resonances with other quantum numbers can also
account for the observed structure. From this discussion it may
already become obvious that the identification of ``missing
resonances'' and their properties from strangeness photoproduction
data, is a heavily debated topic.  We have investigated in how far the
inclusion of a ``missing'' $D_{13}(1895)$ resonance improves the fits
of the $\Sigma$ photoproduction data.  Including the core set of three
$N^*$'s and the $D_{13}(1895)$ in the $s$-channel, we arrive at a best
fit with $\chi^2$ = 5.29. Despite the fact that the $D_{13}$ and
$\Delta ^*$ resonances have their poles in the same energy region,
inspection of Table~\ref{tab:chi} learns that this quality of
agreement is inferior to what was obtained in the calculation with two
$\Delta^*$ resonances. A resonance set consisting of the core of three
$N^*$ resonances, the $D_{13}$ and the two $\Delta^*$'s leads to a fit
with $\chi^2$ = 2.88. Compared to the $\chi^2$ of 3.20, achieved
without introducing the $D_{13}$, this represents only a minor
improvement, in view of the fact that the introduction of a spin-3/2
resonance comes at the expense of throwing in five additional free
parameters in the fitting procedure.

Summarizing the findings of Table~\ref{tab:chi}, we are tempted to
conclude that $\Delta^*$ resonances seem to constitute an essential
part of the dynamics of $\Sigma$ photoproduction.  No convincing
evidence for a salient role for the $D_{13}(1895)$ resonance in
$\Sigma$ photoproduction is found. In this subsection, we have drawn
our conclusions on the basis of numerical calculations within one
particular model (``model D'') for treating the background
diagrams. Alternative models for implementing the background diagrams
will be introduced in the forthcoming section. Anticipating these
investigations, the relative role of the different $N^*$ and
$\Delta^*$ particles turns out to be rather independent of the choices
made with respect to the treatment of the non-resonant diagrams. As it
happens, this will turn out not always to be the case for the
extracted quantitative resonance information.

\subsection{\label{sec:back_cont} Background contributions}

As alluded to in Sec.~\ref{sec:born}, one of the long-standing issues
in modeling strangeness photoproduction is the unrealistically large
amounts of strength produced by the ``bare'' Born terms.  In the
process of trying to counterbalance the strength from these amplitudes
by adding extra ingredients to the theory, it appears that some model
dependence in the treatment of the background terms cannot be avoided.
We now discuss four models which all succeed in cutting down the
background strength in $\Sigma$ photoproduction. In all schemes, the
background contains at least the usual Born terms (including the
``extended'' diagram in the $p(\gamma,K^+)\Sigma^0$ case) and the
$K^*(892)$ vector-meson exchange in the $t$-channel. For the
investigations presented in this subsection, the resonant part
includes the $N^*$ resonances $S_{11}(1650)$, $P_{11}(1710)$ and
$P_{13}(1720)$ and the $\Delta^*$ states $S_{31}(1900)$ and
$P_{31}(1910)$. Those five resonances were identified in
Sec.~\ref{sec:res_contr} as an appropriate set for describing $\Sigma$
photoproduction with a minimal number of free parameters.

\begin{itemize}
\item {\bf Model A:} The hadronic form factors $F_x(\Lambda)$,
described in Eq.~(\ref{eq:formfac}), cut the high-momentum dependence
of the different amplitudes and emerge as a mechanism to reduce the
strength stemming from the Born diagrams to magnitudes of the order of
the measured cross sections.  To fully exploit the power of this
reduction mechanism, we imposed an under limit of 0.4 GeV for the
cutoff mass $\Lambda$ during the fit. Thereby, no $Y^*$ contributions
in the $u$-channel are considered. Despite our reservations regarding
the use of ``soft'' cutoff masses, eventually we arrive in this scheme
at a very satisfactory $\chi^2$ = 2.03 with a cutoff mass $\Lambda$
close to the under limit of 0.4 GeV. It should be stressed that with
cutoff masses as small as the kaon mass, the hadronic form factor
starts playing a predominant role in the description of the reaction
dynamics and heavily affects the predicted values of the observables,
not only in the high-energy regime but even at threshold.
\item {\bf Model B:} In Ref.~\cite{Janssen_role_hyp}, we pointed out that for
the description of $p(\gamma,K^+)\Lambda$ processes, the introduction
of hyperon resonances in the $u$-channel can be an efficient and
physically relevant way of counterbalancing the strength produced by
the Born terms.
\begin{figure}
\resizebox{0.45\textwidth}{!}{\includegraphics{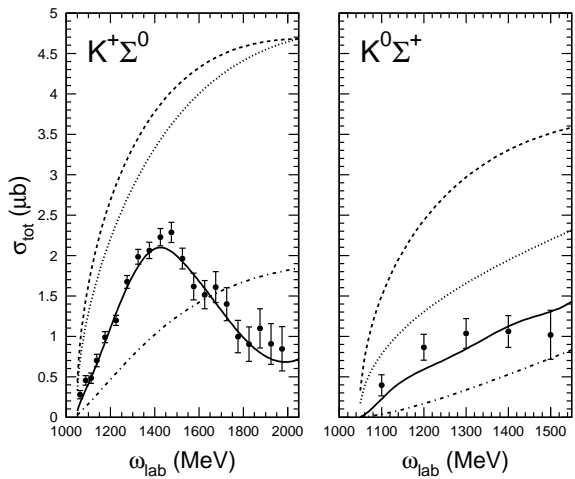}}
\caption{\label{fig:hyp_red} Total cross sections for the $p(\gamma,
K^+)\Sigma^0$ and $p(\gamma,K^0)\Sigma^+$ processes. The dashed curve
denotes the computed strength from the Born terms (with ``strong''
hadronic form factors, $ \Lambda$ = 1.6 GeV). For the dotted line, the
$K^*$ $t$-channel contribution is added. The dot-dashed curve includes
the $Y^*$ hyperon resonances and consequently is the result of the
full background contribution as computed within model B. The solid
line embodies, in addition to the background, the $s$-channel $N^*$
and $\Delta^*$ resonances.  The data are from
Refs.~\protect\cite{Tran,Goers}.}
\end{figure}
More specifically, the destructive interference between the
$u$-channel amplitudes of the $\Lambda^* (1800)$ and $\Lambda^*(1810)$
hyperon resonances and the Born terms resulted in a very satisfactory
description of the $p(\gamma, K^+)\Lambda$ data.  We have made an
attempt to identify an equivalent procedure for $K \Sigma$
photoproduction.  Unfortunately, there is relatively little
theoretical guidance on how to select the proper intermediate hyperon
resonances and how to determine realistic values for their coupling
constants.  Nevertheless, after including the $\Lambda^*(1810)$ and
$\Sigma^*(1880)$ in the $u$-channel, we arrive at a fair description
of the $p(\gamma,K^+)\Sigma^0$ and $p(\gamma,K^0)\Sigma^+$ data with a
$\chi^2$ of 1.95.

Note that the $\Lambda^*$ resonance does not feed the $\gamma p
\rightarrow K^0 \Sigma ^+ $ channel. Consequently, the procedure of
introducing hyperon resonances in the $u$-channel, as a natural
physical mechanism to counteract the background amplitudes, is
expected to be less effective in the $p(\gamma,K^0)\Sigma^+$
channel. However, as can be seen in Fig.~\ref{fig:hyp_red}, a stronger
destructive interference between the $K^*$ vector-meson contribution
in the $t$-channel and the background diagrams is noted for the
$p(\gamma,K^0)\Sigma^+$ process.  One may wonder why this mechanism
does not seem to prevail so strongly in the $\Sigma^0$ photoproduction
case. This can be naturally explained by looking at the respective
electromagnetic coupling constants of the $K^*$ vector mesons.  On the
basis of Eq.~(\ref{eq:K*+K*0}), one finds that the loss of destructive
interference with the $\Lambda^*$ resonance in the
$p(\gamma,K^0)\Sigma^+$ process is likely to be counterbalanced by an
enhanced destructive interference with the $t$-channel vector-meson
exchanges.

One of the obvious advantages of the ``model B'' described here, is
that the role of the hadronic form factors can be diminished to levels
that appear physically acceptable.  Indeed, good fits ($\chi ^2$ =
1.95) of the $\Sigma$ photoproduction data can be obtained with a
``hard'' cutoff mass of the order $\Lambda$ = 1.6~GeV. One argument
that may speak against model B is that the extracted values of the
$Y^*$ coupling constants turn out to be large in comparison with the
$N^*$ and $\Delta^*$ coupling strengths. However, the two $u$-channel
particles, introduced in the computations, could be interpreted as
representing effective particles which account for a larger set of
$u$-channel processes \cite{Janssen_backgr}.
\item {\bf Model C:} A third option is to simply disregard the
constraints of Eq.~(\ref{eq:su3rel}) imposed by (broken) SU(3)-flavor
symmetry. Then, the $g_{K^+ \Lambda p}$ and $g_{K^+ \Sigma^0 p}$
coupling constants can be treated as free parameters in the
minimization procedure. In calculations with model C we are solely
constraining the relative sign between the two coupling constants.
Ignoring $Y^*$ exchange in the $u$-channel, we arrive at an overall
agreement with the data of $\chi^2$ = 1.96 with $g_{K^+ \Lambda
p}/\sqrt{4 \pi} = -0.23$ and $g_{K^+ \Sigma^0 p}/ \sqrt{4 \pi}$ =
0.28.  These numbers are dramatically smaller than what is predicted
on the basis of SU(3)-flavor symmetry (-3.75 and 1.09, respectively).
In this fit, the cutoff mass was allowed to vary freely and adopts a
value of 2.5~GeV.  This value of $\Lambda$ alludes to a rather modest
role for the hadronic form factors in the description of the reaction
dynamics.

\item {\bf Model D:} This scheme is an attempt to unite some of the
virtues of the three models presented above, at the same time
minimizing the number of free parameters that are introduced to
compute the background diagrams.  In this model, the constraints of
Eq.~(\ref{eq:su3rel}) are respected during the fitting procedure.  In
an attempt to keep the model as simple as possible, no $Y^*$ particles
in the $u$-channel are introduced. The hadronic cutoff mass $\Lambda$
is treated as a parameter and allowed to vary freely in a range
defined by the under limit 1.1~GeV.  In the optimum fit, the value of
$\Lambda$ always approaches this under limit, stressing the essential
role of the hadronic form factors for keeping the strength from the
Born diagrams at realistic levels. In Sec.~\ref{sec:res_contr} this
scheme for treating the background diagrams was adopted when
investigating the dominant resonance contributions. This choice was
made on the basis of a minimized number of free parameters related to
the background diagrams. Inspecting Table~\ref{tab:chi}, it becomes
obvious, though, that for a fixed set of resonances model D
systematically leads to $\chi^2$ values which are inferior to those
obtained in models A, B and C.

\end{itemize}

\begin{figure}
\resizebox{0.45\textwidth}{!}{\includegraphics{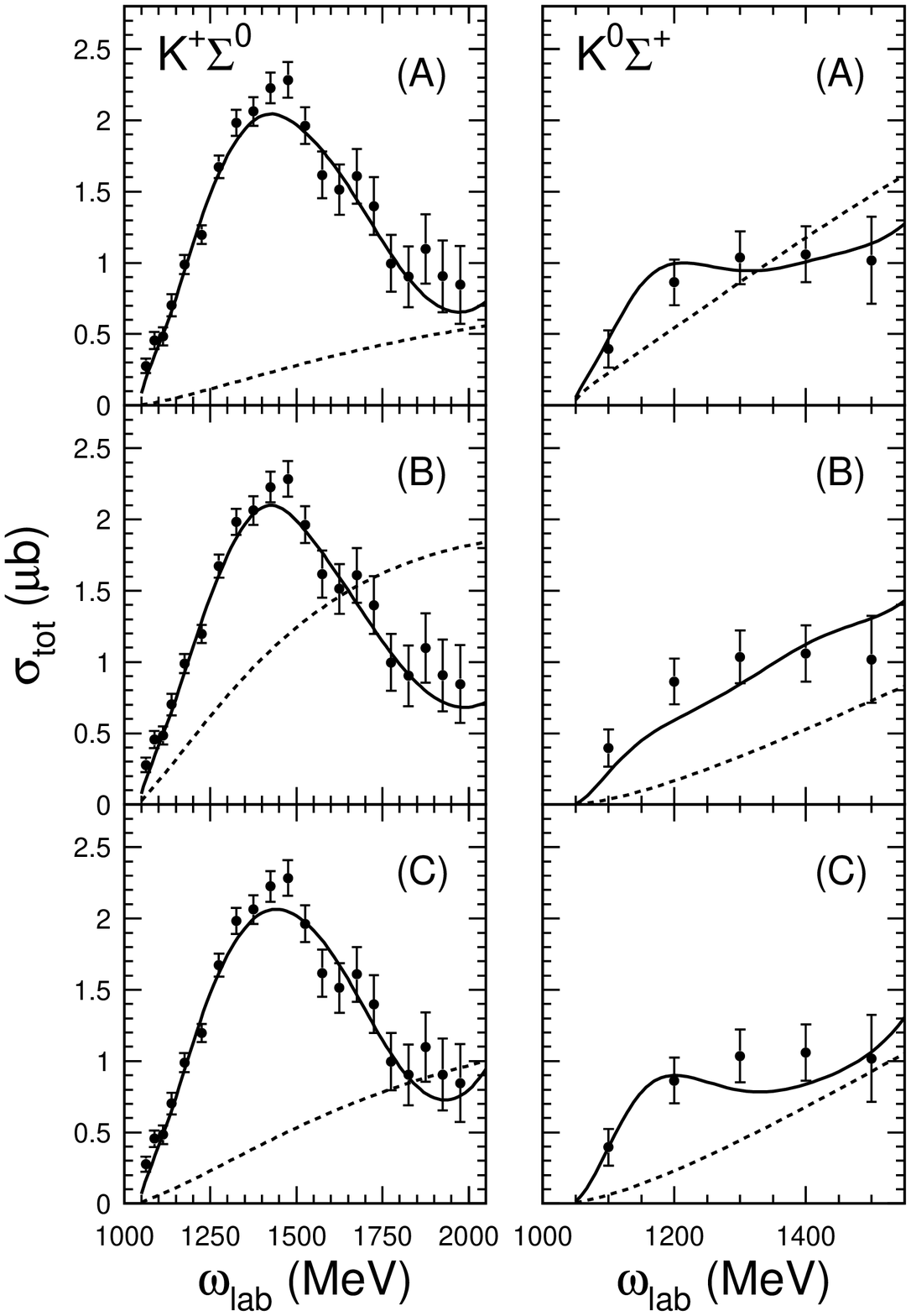}}
\caption{\label{fig:3_back} Energy dependence of the total
$p(\gamma, K^+)\Sigma^0$ and $p(\gamma,K^0)\Sigma^+$ cross
sections. The dashed curves denote the computed strength from the
``background'' diagrams. The solid curves include both the background
and the resonant amplitudes. The panels (A), (B), (C) refer to the
optimum fits obtained with the three background models A, B and C,
respectively.  The data are from Refs.~\protect\cite{Tran,Goers}.}
\end{figure}

\begin{table*}
\begin{center}
%\resizebox{\columnwidth}{!}{
\begin{tabular}{||l|c||r|r|r||}
\hline \hline
 & & & & \\
 & & \multicolumn{1}{|c|}{model A} & \multicolumn{1}{|c|}{model B} &
 \multicolumn{1}{|c||}{model C} \\ 
 & & & & \\
\hline
Born terms & $g_{\sss K^+ \Sigma p} / \sqrt{4 \pi}$ & 9.949$\cdot$10$^{-1}$ &
 9.215$\cdot$10$^{-1}$ &  2.811$\cdot$10$^{-1}$ \\

     & $g_{\sss K^+ \Lambda p} / \sqrt{4 \pi}$ &
-4.487\phantom{$\cdot$10$^{-0}$}  &
     -3.017\phantom{$\cdot$10$^{-0}$} &  -2.261$\cdot$10$^{-1}$ \\

$K^*$ & $G^v_{K^*}$ & 6.926$\cdot$10$^{-2}$ &  7.598$\cdot$10$^{-2}$ &
6.828$\cdot$10$^{-2}$ \\ 

      & $G^t_{K^*}$ & 8.047$\cdot$10$^{-2}$ &  4.087$\cdot$10$^{-2}$ &
      1.411$\cdot$10$^{-1}$ \\ 

$P_{01}(1810)$ & $G_{P_{01}}$ & &  21.838\phantom{$\cdot$10$^{-0}$} & \\

$P_{11}(1880)$ & $G_{P_{11}}$ & &  -9.451\phantom{$\cdot$10$^{-0}$} & \\
\hline
$S_{11}(1650)$ & $G_{S_{11}}$ & -4.568$\cdot$10$^{-2}$ &
-4.516$\cdot$10$^{-3}$ & -2.511$\cdot$10$^{-2}$ \\

$P_{11}(1710)$ & $G_{P_{11}}$ & -1.213$\cdot$10$^{-1}$ &
-1.583$\cdot$10$^{-1}$ & -1.879$\cdot$10$^{-1}$ \\ 

$P_{13}(1720)$ & $G^{\left(1 \right)}_{P_{13}}$ &  2.367$\cdot$10$^{-2}$ &
1.706$\cdot$10$^{-2}$ &  2.699$\cdot$10$^{-2}$ \\

               & $G^{\left(2 \right)}_{P_{13}}$ &  5.238$\cdot$10$^{-2}$ &
8.343$\cdot$10$^{-2}$ &  5.213$\cdot$10$^{-2}$ \\

               & $X_{P_{13}} $ &  12.351\phantom{$\cdot$10$^{-0}$} &
               6.943\phantom{$\cdot$10$^{-0}$} &
               14.863\phantom{$\cdot$10$^{-0}$}  \\ 

               & $Y_{P_{13}} $ &  3.781\phantom{$\cdot$10$^{-0}$} &
               4.765\phantom{$\cdot$10$^{-0}$} &
               3.861\phantom{$\cdot$10$^{-0}$} \\ 

               & $Z_{P_{13}} $ &  -1.122\phantom{$\cdot$10$^{-0}$} &
               -1.129\phantom{$\cdot$10$^{-0}$} &
               -1.089\phantom{$\cdot$10$^{-0}$}  \\

$S_{31}(1900)$ & $G_{S_{31}}$ &  5.131$\cdot$10$^{-2}$ &
4.279$\cdot$10$^{-2}$ & 4.351$\cdot$10$^{-2}$ \\

$P_{31}(1910)$ & $G_{P_{31}}$ &  3.726$\cdot$10$^{-1}$ &
3.599$\cdot$10$^{-1}$ & 3.920$\cdot$10$^{-1}$ \\ 
\hline
cutoff mass  & $\Lambda_{born}$ & 439.68\phantom{$^{-2}$} &
1605.04\phantom{$^{-2}$} &  2509.22\phantom{$^{-2}$}  \\

      & $\Lambda_{res}$ & 1616.20\phantom{$^{-2}$} &
1602.43\phantom{$^{-2}$} & 1601.54\phantom{$^{-2}$} \\ 
\hline 

\multicolumn{1}{||c|}{$\chi^2$} & & \multicolumn{1}{|c|}{2.03}
& \multicolumn{1}{|c|}{1.95} & \multicolumn{1}{|c||}{1.96} \\
\hline \hline 
\end{tabular}
%}
\caption{\label{tab:num_cc_ks}Numerical values of the coupling
constants and hadronic cutoff masses (in MeV) in the isobar model
calculations for the $p\left(\gamma, K^+\right) \Sigma^0$ process. The
conventions for the coupling constants are as in
Ref.~\protect\cite{Janssen_backgr}.}
\end{center}
\end{table*}

With all suggested models for implementing the background diagrams, we
arrive at a satisfactory description of the available data. This
feature becomes apparent from the $\chi^2$ values contained in
Table~\ref{tab:chi} and can also be illustrated by directly comparing
model predictions with the data points. Fig.~\ref{fig:3_back} shows
the energy dependence of the total cross section for the background
models A, B and C. In these plots, the strength from the background
diagrams is also shown.  It is clear from this figure that the
predictions for the background and resonant strength are qualitatively
similar for models A and C.  Nevertheless, there are major differences
between the assumptions underlying the two models.  Most importantly,
whereas model A is based on $g_{KYp}$ coupling constants respecting
(broken) SU(3) flavor symmetry, model C does not impose any
constraints of that type. However, it appears that respecting SU(3)
flavor symmetry comes at a certain price.  Indeed, when adopting SU(3)
constraints on the $g_{KYp}$ values, one appears to be forced to
either introduce (unrealistically) soft hadronic form factors (model
A), or implement additional $u$-channel diagrams (model B) to
destructively interfere with the Born terms. In the literature on
$p(\gamma,K)Y$, there is no agreement on whether or not to respect
SU(3) symmetry. The analysis of Lee {\em et al.} in
Ref.~\cite{Lee_nuclei} is based on moderately broken SU(3) symmetry
($g_{K\Lambda p}/\sqrt{4 \pi}$ = -3.80 and $g_{K\Sigma p}/\sqrt{4
\pi}$ = 1.20). This for example also holds for the work of David {\em
et al.}  \cite{David} ($g_{K\Lambda p}/\sqrt{4 \pi}$ = -3.23 and
$g_{K\Sigma p}/\sqrt{4 \pi}$ = 0.80). Other analyses yielded $g_{KYp}$
coupling constants which do not respect SU(3) symmetry. Williams et
al. \cite{Williams} were one of the first to document that by not
imposing flavor symmetry a fair description is obtained with much
smaller couplings, -2.38 $\leq g_{K \Lambda p}/ \sqrt{4 \pi} \leq$ -1.16
and 0.0928 $\leq g_{K \Sigma p}/\sqrt{4 \pi} \leq$ 0.273. Similarly
and more recently, Mart {\em et al.}  \cite{Mart_sigma_1} use
$g_{K\Lambda p}/\sqrt{4 \pi}$ = 0.51 and $g_{K\Sigma p}/\sqrt{4 \pi}$
= 0.13, Feuster and Mosel \cite{Feuster2} extract $g_{K \Lambda p} /
\sqrt{4 \pi}$ = -1.72 and Hsiao {\em et al.} \cite{Hsiao} arrives at
values ranging between -2.41 $\leq g_{K\Lambda p}/\sqrt{4 \pi} \leq$
-1.24 and -0.50 $\leq g_{K\Sigma p}/\sqrt{4 \pi} \leq$ 1.04. All these
absolute $g_{KYp}$'s are substantially smaller than what could be
expected on the basis of moderately broken SU(3) flavor symmetry. The
numerical values obtained in this work are summarized in
Table~\ref{tab:num_cc_ks}. As pointed out in Ref.~\cite{Feuster2}, it
probably makes more sense to compare the product $g_{KYp} \cdot
\widehat{F}$ than the bare coupling constants $g_{KYp}$. In this
respect, our model A, based on SU(3) flavor symmetry for $g_{KYp}$,
requires form factors $\widehat{F}$ of the order 0.1. Model C, on the
other hand, use values of $\widehat{F}$ of the order 1, but the
$g_{KYp}$ are close to one-tenth of the SU(3) predictions. This
results in comparable values for $g_{KYp} \cdot \widehat{F}$ in both
models. In model B, SU(3) flavor symmetry is only mildly violated and
due to the larger cutoff mass, the product $g_{KYp} \cdot \widehat{F}$
is considerably larger than in the models A and C.

\begin{figure*}
\resizebox{0.95\textwidth}{!}{\includegraphics{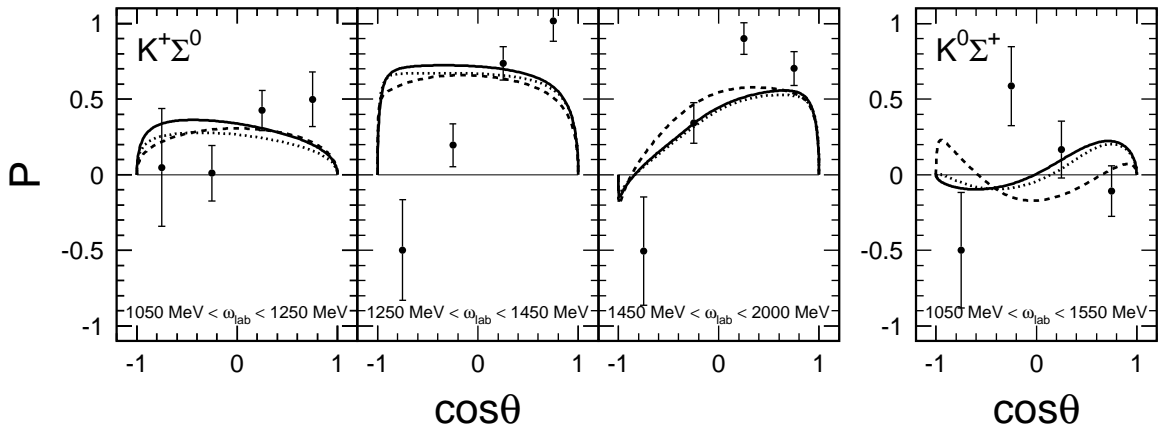}}
\caption{\label{fig:recpol} Angular dependence of the $p(\gamma,
K^+)\vec{\Sigma}^0$ and $p(\gamma, K^0)\vec{\Sigma}^+$
recoil-polarization asymmetry (P). The solid, dashed and dotted lines
are obtained with background model A, B and C, respectively. Our
results are averaged over the experimental energy bins.  The data are
from Refs.~\protect\cite{Tran,Goers}.}
\end{figure*}

In Fig.~\ref{fig:recpol}, model calculations for the angular
distribution of the $p(\gamma,K^+)\vec{\Sigma}^0$ and
$p(\gamma,K^0)\vec{\Sigma}^+$ recoil-polarization asymmetry are
given. This asymmetry is defined as:
\begin{equation}
P = \frac{ d \sigma / d \Omega^{\left( + \right)} - d \sigma / d
\Omega^{\left( - \right)} }{ d \sigma / d \Omega^{\left( + \right)} + d
\sigma / d \Omega^{\left( - \right)} } \;,
\label{eq:rec_asym}
\end{equation}
where $+ \left(- \right)$ refers to a hyperon polarization parallel
(anti-parallel) to the $\left(\vec{p}_\gamma \times \vec{p}_K
\right)$-axis.  Note, however, that the data points for the asymmetry
of the $K^0 \Sigma^+$ process, are binned over the whole energy
range. Consequently, they hardly affect the $\chi^2$ of the global
fit.

\begin{figure*}
\resizebox{0.75\textwidth}{!}{\includegraphics{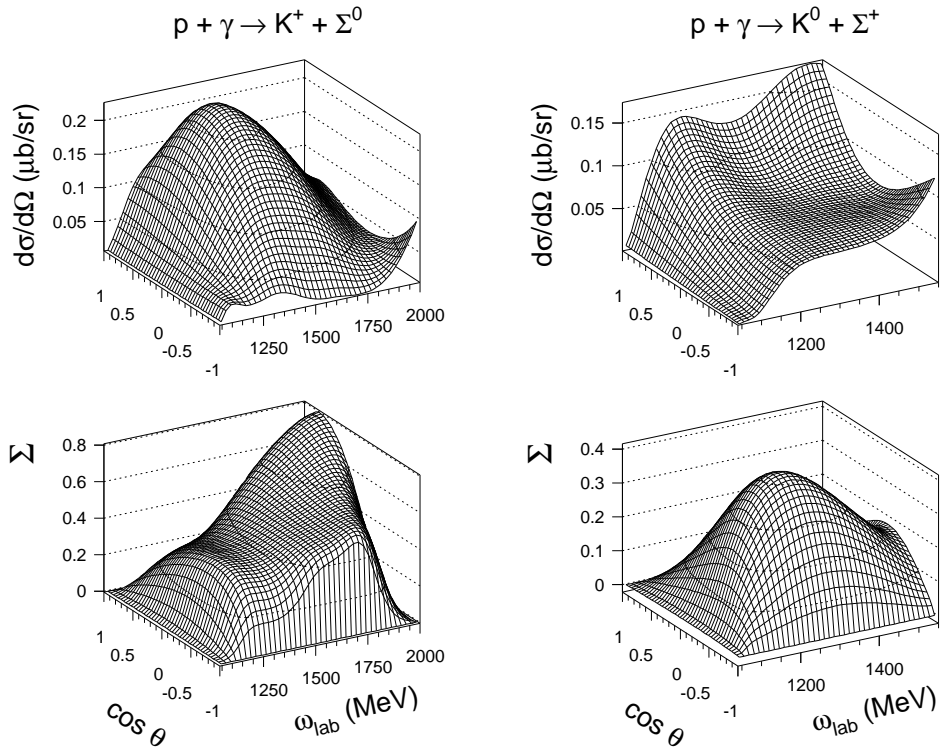}}
\caption{\label{fig:diffcs.a.2dim}The differential cross section and
photon-beam asymmetry ($\Sigma$) for the $p(\gamma,K^+)\Sigma^0$ and
$p(\gamma,K^0)\Sigma^+$ processes as a function of the photon lab energy
$\omega_{lab}$ and $\cos \theta$.  The calculations account for the
$S_{11}(1650)$, $P_{11}(1710)$, $P_{13}(1720)$, $S_{31}(1900)$ and
$P_{31}(1910)$ $s$-channel resonances.
The background terms are treated according to the prescriptions of model A.}
\end{figure*} 
\begin{figure*}
\resizebox{0.75\textwidth}{!}{\includegraphics{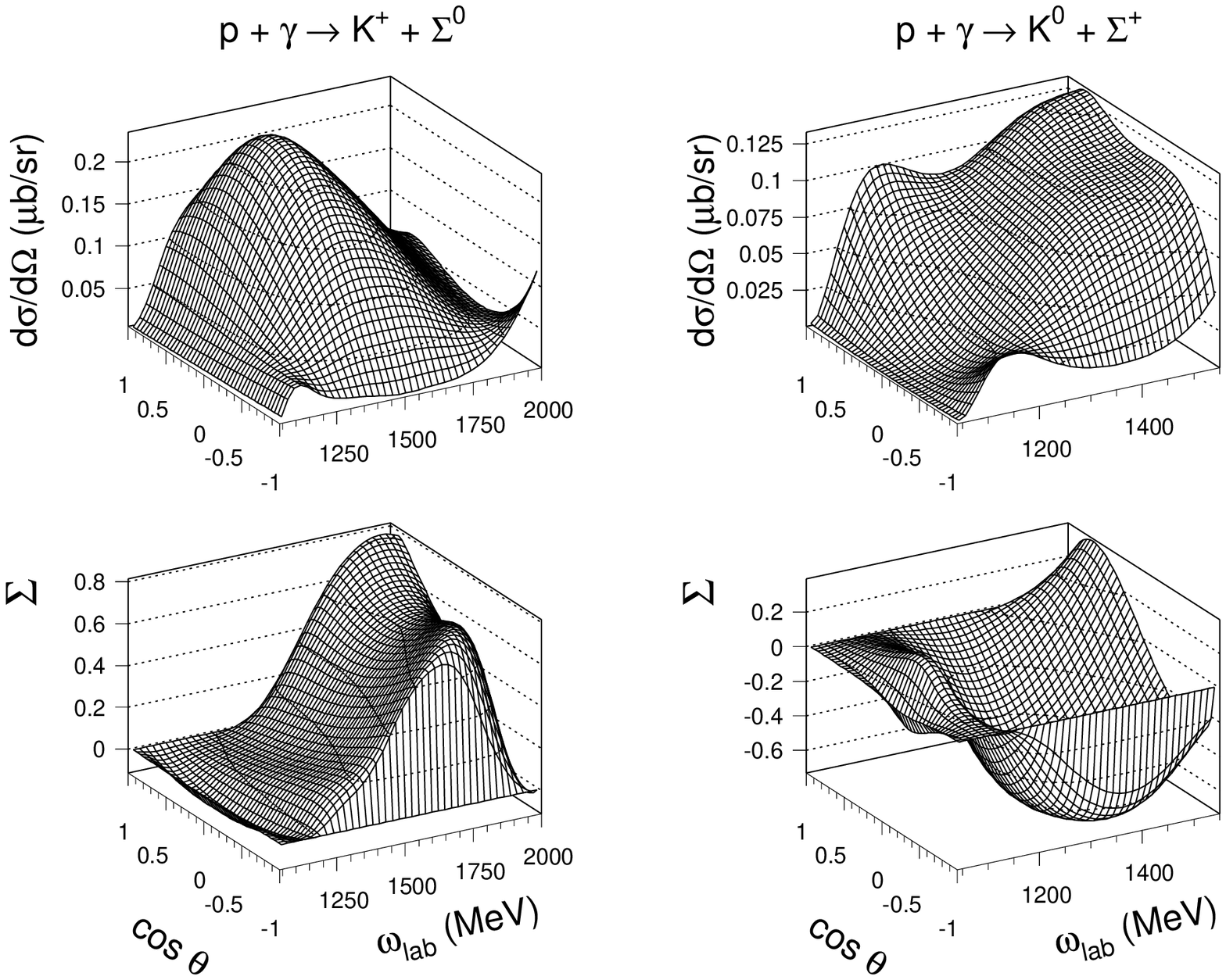}}
\caption{\label{fig:diffcs.b.2dim}As in
Figure~\ref{fig:diffcs.a.2dim} but now for background model B.}
\end{figure*} 
\begin{figure*}
\resizebox{0.75\textwidth}{!}{\includegraphics{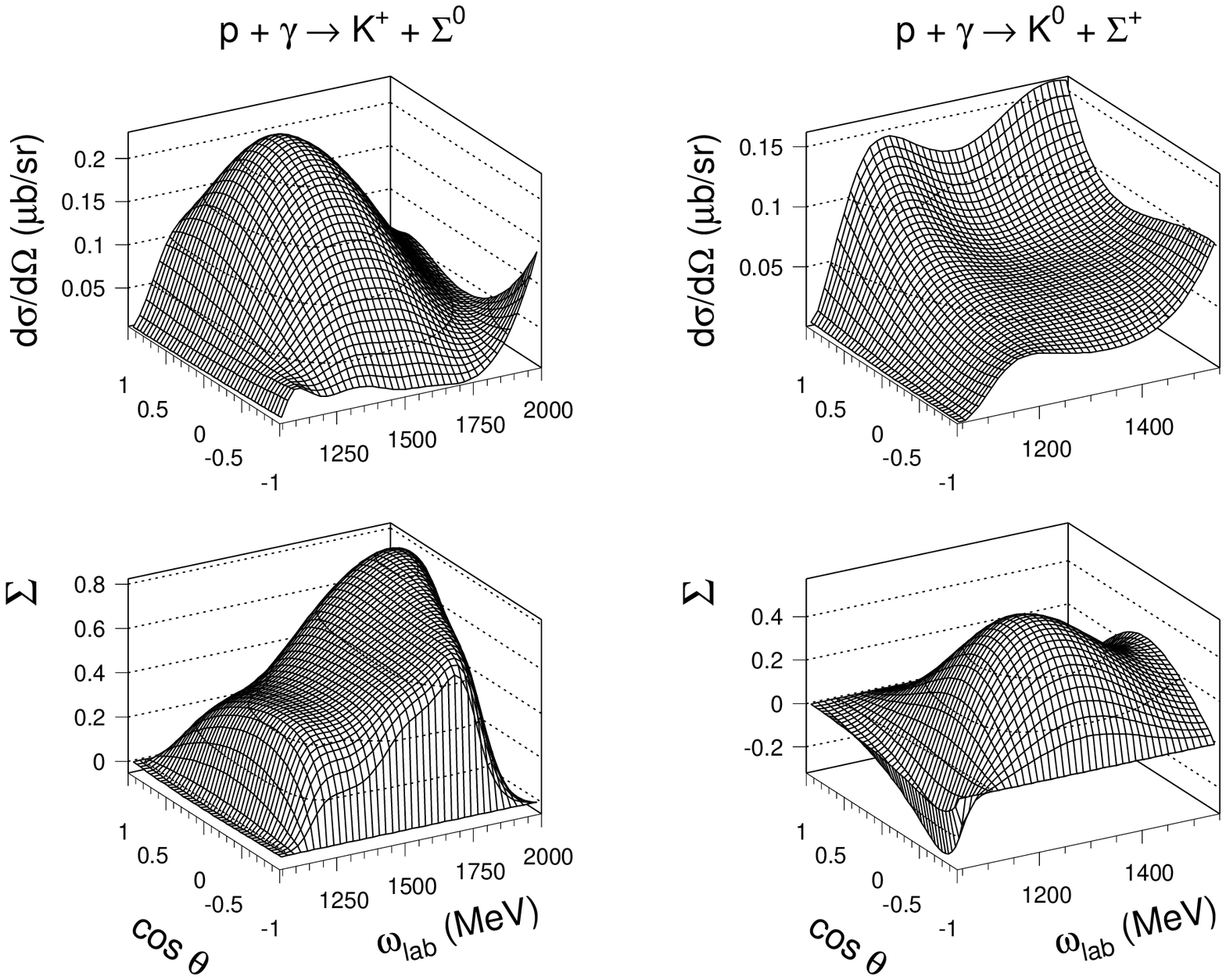}}
\caption{\label{fig:diffcs.c.2dim}As in
Figure~\ref{fig:diffcs.a.2dim} but now for background model C.}
\end{figure*} 

\begin{figure*}
\resizebox{0.75\textwidth}{!}{\includegraphics{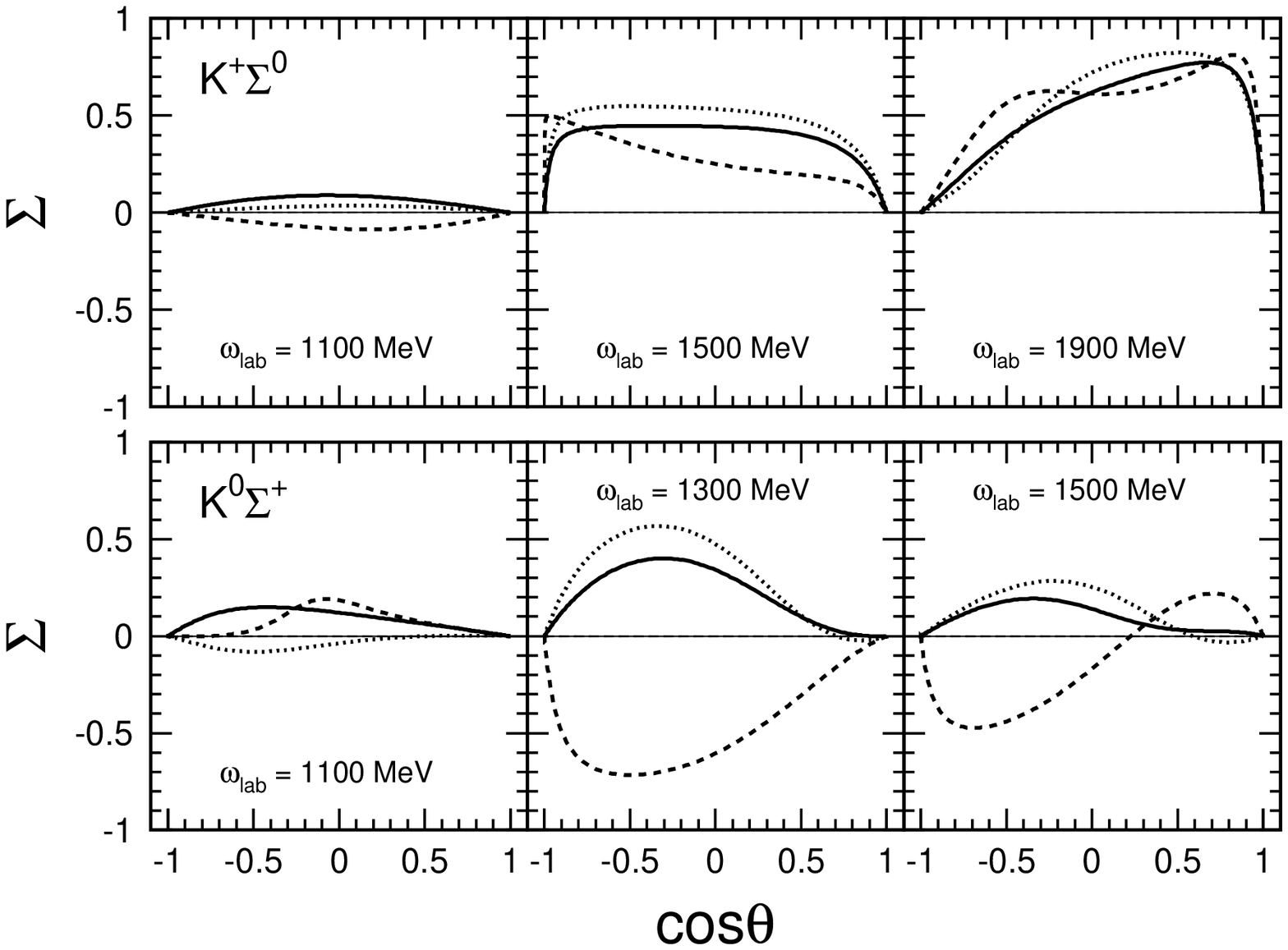}}
\caption{\label{fig:phopol} The angular dependence of the photon-beam
asymmetry ($\Sigma$) for three photon lab energies. The solid, dashed
and dotted curves are the predictions as obtained with the background
model A, B and C, respectively. The upper panels are for the
$p(\vec{\gamma}, K^+)\Sigma^0$ process, the lower for the
$p(\vec{\gamma}, K^0)\Sigma^+$ reaction.}
\end{figure*}

Predictions for the energy and angular dependence of the differential
cross section and the photon-beam asymmetry ($\Sigma$) are displayed in
Figs.~\ref{fig:diffcs.a.2dim}, \ref{fig:diffcs.b.2dim} and
\ref{fig:diffcs.c.2dim} for the models A, B and C described above.
The photon-beam asymmetry is defined in the standard manner:
\begin{equation}
\Sigma = \frac{ d \sigma / d \Omega^{\left( \perp \right)} - d
\sigma / d \Omega^{\left( \parallel \right)} }{ d \sigma / d
\Omega^{\left( \perp \right)} + d \sigma / d \Omega^{\left( \parallel
\right)} } \;. \label{eq:pho_asym}
\end{equation}
Here, $\perp \left(\parallel  \right)$ refers to linearly polarized
photons perpendicular (parallel) to the reaction plane.
From visual inspection of the
Figs.~\ref{fig:diffcs.a.2dim}-\ref{fig:diffcs.c.2dim}, it indeed
becomes apparent that the energy and angular dependence of the
differential cross sections is rather similar for the three
models. The sudden rise in the predicted $\Sigma ^0$ cross sections at
very backward angles and the highest photon energies should not be
considered as physical.  It illustrates the limits of the hadronic
models for predicting observables in ``unmeasured'' regions of the
phase space. Note that the data used in the fitting procedure do not
extend beyond 2.0 GeV ($\Sigma^0$ production) and 1.55 GeV ($\Sigma^+$
production).  For the angular and energy dependence of the
$p(\vec{\gamma},K^+)\Sigma^0$ photon-beam asymmetry, models A, B and C
produce comparable results. Although no published data exist for this
observable to date, the model dependences in the predictions for this
observable seem to be modest. On the other hand, large variations
between the different predictions for the $p(\gamma,K^0)\Sigma^+$
photon-beam asymmetry are observed.  To fully appreciate this, we have
gathered the calculations for the photon-beam asymmetries at some
fixed photon lab energies in Fig.~\ref{fig:phopol}.
With no doubt, more precise data for the various polarization
observables would help in further constraining the model dependences
in the treatment of the background diagrams.

Not only the predictions for some of the asymmetries, but also the
extraction of resonance parameters, turns out to be reasonably
sensitive to the adopted procedure to treat the background.
\begin{figure*}
\resizebox{0.8\textwidth}{!}{\includegraphics{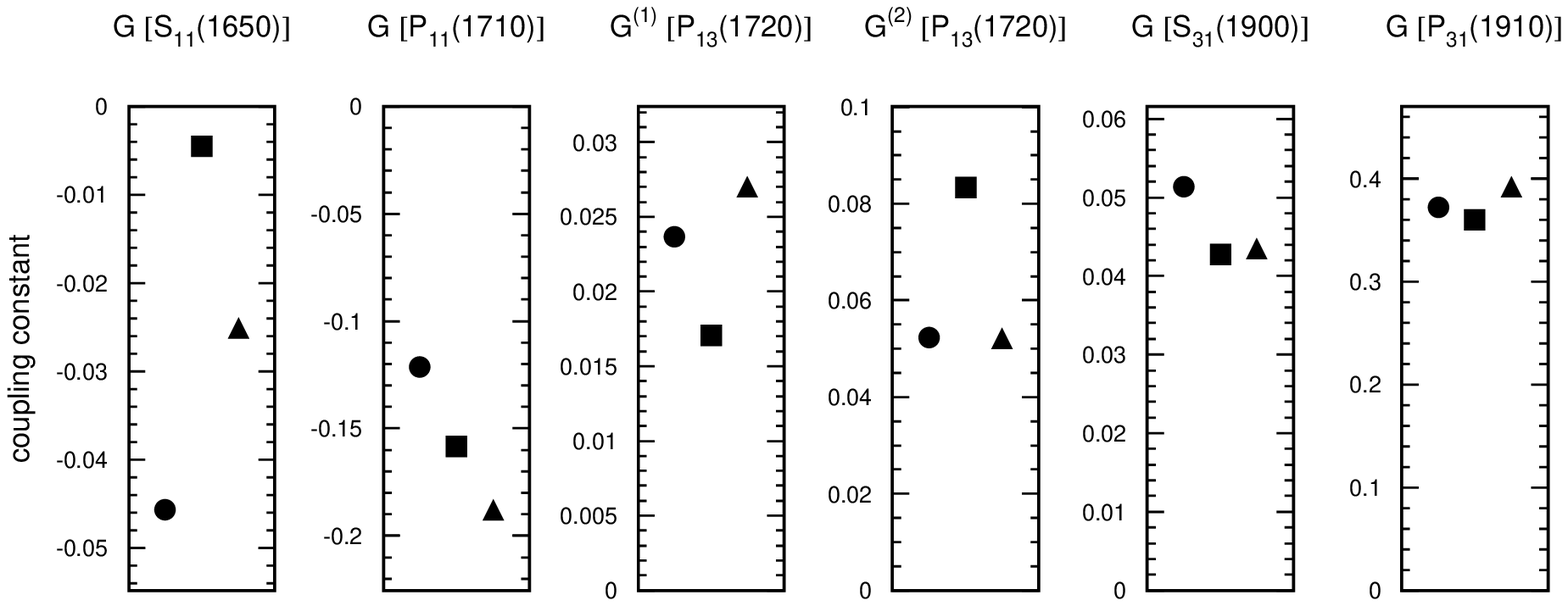}}
\caption{\label{fig:cc_stab} The extracted coupling constants for the
$N^*$ and $\Delta^*$ resonances. The circles are extracted within
model A, the squares within model B and the triangles within model
C. The conventions adopted for the coupling constants are summarized
in Ref.~\cite{Janssen_backgr}.}
\end{figure*}
This feature is illustrated in Fig.~\ref{fig:cc_stab}, where the
extracted resonance coupling constants are plotted for the background
models A, B and C. For the $I = \frac{3}{2}$ $\Delta^*$ resonances,
the extracted coupling constants are rather insensitive to the choices
for the background. Larger variations, exceeding the 20\% level are
observed for the extracted $N^*$ parameters. Table~\ref{tab:num_cc_ks}
lists the numerical values of all parameters corresponding with the
optimum fits. The discerned model variation in the resonance
parameters show that a model-independent extraction of this
information from the strangeness-photoproduction data is not yet at
hand. This is rather unfortunate, given that these variables play a
crucial role in linking the predictions of (constituent) quark models
and the photoproduction data.

To conclude this Section, we come back to the aforementioned issue of
the missing $D_{13}$ nucleon resonance. Adopting background model D,
the inclusion of this $N^*$ particle improved the quality of the
global fit from $\chi^2$ = 3.20 to $\chi^2$ = 2.88. We stress again
that this comes at the expense of adding five extra parameters. We
have investigated whether a similar qualitative feature emerged with
background models A, B and C. In all cases, a global fit with the core
of the three $N^*$ and the two $\Delta^*$ resonances with and without
the $D_{13}(1895)$ was performed. The results are contained in
Table~\ref{tab:chi}. The improvement in the quality of the fit varied
from 3\% (model A) to 8\% (model B). We appreciate this improvement as
rather modest in view of the five extra parameters. For the sake of
reference, we mention that in a comparable tree-level analysis of the
$p(\gamma,K^+)\Lambda$ data, the inclusion of an extra $D_{13}(1895)$
resonance improved the fits from 22\% up to 40\%, depending on the
choices made for computing the background contributions
\cite{Janssen_role_hyp}.

\subsection{\label{sec:radcap} Radiative Kaon Capture}

Through the rules of crossing symmetry \cite{Ji}, the process of
radiative kaon capture:
\begin{equation}
K^- + p \rightarrow \gamma + \Sigma^0 \;,
\end{equation}
is related to the kaon photoproduction $p(\gamma, K^+)\Sigma^0$
process. Indeed, one has:
\begin{eqnarray}
\lefteqn{{\cal M}^{K^- p \rightarrow \gamma \Sigma^0} \left(p, k,
p_{\sss K}, p_{\sss \Sigma} \right) =} \nonumber \\ 
& &{\cal M}^{\gamma p \rightarrow K^+ \Sigma^0} \left(p, - k, - p_{\sss K},
p_{\sss \Sigma} \right) \;,
\end{eqnarray}
where $p$, $k$, $p_{\sss K}$ and $p_{\sss \Sigma}$ are the four
momenta of the proton, photon, kaon and $\Sigma$,
respectively. Unfortunately, to our knowledge the sole reliable data
point for the radiative kaon capture process is for the branching
ratio of {\em stopped} kaons:
\begin{eqnarray}
{\cal R} &=& \frac{ \Gamma \left(K^- p \rightarrow \gamma \Sigma^0
\right) }{\Gamma \left( K^- p \rightarrow \mbox{all} \right) }
\nonumber \\ 
&=& \frac{\pi}{2 W_{K^-p}} \frac{\omega}{M_p M_{\sss K}
\left(M_p + M_{\sss K} \right)} \frac{1}{\left(4 \pi \right)^2}
\nonumber \\
& & \times \frac{1}{2} \left| {\cal M}^{\gamma p \rightarrow K^+
\Sigma^0} \left(p, - k, - p_{\sss K}, p_{\sss Y} \right) \right|^2 \;, 
\end{eqnarray}
where $\omega$ is the c.m. photon energy and $W_{K^-p}$ = 560 $\pm$
135 MeV fm$^3$ is the $K^-p$ pseudo-potential determined by Burkhardt
{\em et al.}  \cite{Burkhardt}.

\begin{table}
\begin{center}
\resizebox{\columnwidth}{!}{
\begin{tabular}{||c|ccc|c|c||}
\hline \hline 
&  A &  B &  C &  B +  & experiment \\ 
& & & & $\Lambda^*(1405)$ & Ref.~\cite{Whitehouse} \\ 
\hline 
${\cal R} \times 10^3$ & 0.016 & 0.259 & 0.002 & 1.556 & 1.44 $\pm$ 0.20
$\pm$ 0.11 \\ 
\hline \hline
\end{tabular}
}
\caption{\label{tab:radcap} Model predictions with the background
models A, B and C for the branching ratio ${\cal R} = \Gamma \left(K^-
p \rightarrow \gamma \Sigma^0 \right) / \Gamma \left( K^- p
\rightarrow \mbox{all} \right)$. The experimental value is from
Ref.~\cite{Whitehouse}. }
\end{center}
\end{table}

Our model predictions for ${\cal R}$ are summarized in
Table~\ref{tab:radcap}. It becomes clear that the three proposed
background models produce values which differ by two orders of
magnitude and considerably underestimate the measured value. However,
it has been stressed by various authors \cite{Burkhardt,David,Workman}
that the $\Lambda^*(1405)$ resonance, which is an $s$-channel
resonance just below the decay threshold, is of crucial importance for
reproducing ${\cal R}$. In the $\Sigma$ photoproduction process, the
$\Lambda^*(1405)$ is a candidate for a resonant $u$-channel
contribution. However, in our analysis of $\gamma p \rightarrow K
\Sigma$, no direct need for introducing the $\Lambda^*(1405)$ emerged
(model B is the sole model that implements $u$-channel $\Lambda^*$ and
$\Sigma^*$ resonances). Therefore, we investigated whether a
consistent description of the branching ratio ${\cal R}$ and the
$p(\gamma,K)\Sigma$ SAPHIR data is feasible. To this end, we started
out from background model B to which we added the $\Lambda^*(1405)$
hyperon resonance. In the fitting procedure against the SAPHIR data
and the value of ${\cal R}$, we arrive at a $\chi^2$ value of
1.92. The result for ${\cal R}$ = 1.56~$\cdot$~10$^{-3}$ is in
agreement with the experimental value within the error bars.

To conclude this subsection, we wish to stress that the branching
ratio ${\cal R}$ in radiative kaon capture has only a limited
potential to constrain the $p(\gamma,K)\Sigma$ reaction
dynamics. Indeed, there are strong indications that the value of
${\cal R}$ is mainly determined by the strength of the
$\Lambda^*(1405)$ while this resonance plays only a secondary role in
$\gamma p \rightarrow K \Sigma$ reactions.

\section{\label{sec:concl} Conclusion}

In this work, we have presented results for neutral and charged
$\Sigma$ photoproduction off the proton in a hadronic model at tree
level.  By comparing model calculations to the SAPHIR data, we were
able to identify a set of five $N^*$ and $\Delta^*$ resonances
($S_{11}(1650)$, $P_{11}(1710)$, $P_{13}(1720)$, $S_{31}(1900)$ and
$P_{31}(1910)$) with which a satisfactory description of the data can
be obtained.  Our calculations do not provide evidence for a salient
role for the $D_{13}(1895)$ resonance in $\Sigma$ photoproduction. The
$D_{13}(1895)$ has never been observed in $\pi N$ scattering.
Recently, it was claimed that signals for the existence of such a
resonance emerge from the $\Lambda$ photoproduction data.

We have shown that the ``bare'' Born amplitudes produce
$p(\gamma,K)\Sigma$ cross sections which dramatically overshoot the
measured ones. Therefore, additional ingredients in the model
calculations beyond resonance contributions appear essential.  We have
presented total and differential cross sections as well as recoil and
photon asymmetry results for four schemes which accomplish to
cut down satisfactorily the magnitude of the Born amplitudes.  Through
the background diagrams some model dependence in the extracted
resonance parameters gets introduced. This dependence turns out to be
small for the $\Delta^*$ particles. For some of the $N^*$ resonances,
though, the obtained coupling constants may vary up to a factor of
two, depending on which model is used to implement the
background. Predictions for the $p(\gamma, K^+)\Sigma^0$ photon-beam
asymmetry are only moderately sensitive to the implementation of the
background terms. This is not the case for the photon-beam asymmetries
in the $p(\gamma,K^0)\Sigma^+$ channel. It should be stressed that for
the $\Sigma^0$ case far more data are presently available. In the
foreseeable future, the available amount of $\Sigma$ photoproduction
data in the resonance region will dramatically increase.  Research
efforts at Jefferson Lab \cite{Schumacher}, at GRAAL (Grenoble)
\cite{DAngelo} and SPring-8 \cite{Zegers} and continuing analysis work
from the SAPHIR collaboration \cite{Glander} will extend the $\Sigma$
and $\Lambda$ photo and electroproduction data beyond 2.0 GeV and will
shed light on the (in)capability of hadronic approaches to model the
physics at higher photon energies. They will also provide large and
accurate sets of polarization data.  With such an extended data base,
one can be hopeful to better constrain the theoretical models and
reveal the full dynamics of strangeness photoproduction reactions.

\appendix*

\section{\label{sec:cc_rel} Isospin symmetry and coupling constants}

Isospin symmetry considerations are extremely useful tools to
establish ranges and relative signs between series of coupling
constants.  In this Appendix we sketch how isospin arguments can be
used to establish relations between the different hadronic and
electromagnetic coupling constants which are required in global fits
to photoinduced open strangeness production on the proton.  In this
Appendix, we assume the isospin symmetry of the various meson and
baryon multiplets to be exact.  In what follows we will briefly
address both hadronic and electromagnetic coupling constants.

\subsection{\label{sec:had_cc_rel}Hadronic decays of $N^*$ and
$\Delta^*$ resonances} 

The calculation of hadronic transitions of baryon resonances poses a
challenging task to Constituent Quark Models (CQM).  The major
difficulty of such models is to determine the structure of the
operators which govern the decay mechanism.  This reflects the
insufficient basic insight into the quark dynamics in low-energy
hadron phenomenology.  Most CQM's studying hadronic decays of baryon
resonances (for a recent example see Ref.~\cite{Theussl}), start from
a transition operator at quark level which does not contain
isospin-dependent terms. In such a model, the amplitude for a
pseudo-scalar hadronic decay of the type:
\begin{equation}
B(I_1, \; M_1) \longrightarrow K(I_2, \;  M_2) \; \;  + \; \;  Y(I_3, \;
M_3) \; , 
\end{equation} 
is proportional to the isospin part:
\begin{equation}
\frac { (-1)^{I_2-I_1} } {\sqrt{2 I_1 + 1}} \left< I_2 \;
M_2 \; I_3 \; M_3 \mid I_1 \; M_1 \right> 
\left< I_2   \parallel \widehat{T}^{(I_3)} \parallel I_1 \right>,
\label{eq:isospi}
\end{equation} 
where $I_i$ ans $M_i$ are the isospin and isospin projection of the
respective particles and $\widehat{T}^{(I)}$ denotes a spherical
tensor operator of rank $I$.  From the expression (\ref{eq:isospi}),
one easily obtains the following relations between the different
isospin channels in $ N \rightarrow K \Sigma$:
\begin{equation}
g_{K^+ \Sigma^0 p}\ =\ \frac{g_{K^0 \Sigma^+ p}}{\sqrt{2}}\ =\ -\
g_{K^0 \Sigma^0 n}\ =\ \frac{g_{K^+ \Sigma^- n}}{\sqrt{2}} \;.
\label{eq:SKN}  
\end{equation}
In determining these relations we adopted the following conventions
for the isospin states of the {\em physical} $\Sigma$ particles:
\begin{eqnarray}
\Sigma ^+ & \; : \; & - \left|I=1, \; M =+1 \right> \;,\nonumber \\
\Sigma ^0 & \; : \; & + \left|I=1, \; M = 0 \right> \;, \\
\Sigma ^- & \; : \; & + \left|I=1, \; M =-1 \right> \;. \nonumber 
\end{eqnarray}
For the hadronic decays of the type $ N \rightarrow K \Lambda$,
starting from Eq.~(\ref{eq:isospi}), even simpler 
relations can be written down:
\begin{eqnarray}
g_{\Lambda K^+ p } &=& g_{\Lambda K^0 n } \;. 
\label{eq:LKN}
\end{eqnarray}
We now consider hadronic decays of the type $\Delta \rightarrow
K \Sigma$. Defining the corresponding isospin states for the
$\Delta^{+,0}$ particles as $ \left| I = \frac{3}{2},\ I_3 = \pm\
\frac{1}{2} \right>$, one obtains the following relations from
Eq.~(\ref{eq:isospi}):
\begin{eqnarray}
\lefteqn{g_{K^+ \Sigma^0 \Delta^+} = -\ \sqrt{2}\ g_{K^0 \Sigma^+
\Delta^+} = } \nonumber \\
& & g_{K^0 \Sigma^0 \Delta^0} = \sqrt{2}\ g_{K^+ \Sigma^- \Delta^0} \;. 
\label{eq:SKD}
\end{eqnarray}
The relations contained in Eqs.~(\ref{eq:SKN}),~(\ref{eq:LKN}) and
(\ref{eq:SKD}), also hold when a $N^*$, $K^*$, $\Sigma^*$ or
$\Lambda^*$ resonance is involved at the vertex.

\subsection{Electromagnetic vertices}

Most of the information with respect to electromagnetic couplings rely on
experimental quantities. The measured decay widths for the $K^{*+}(892)$ and
$K^{*0}(892)$ vector mesons are \cite{PDG}:
\begin{eqnarray}
\Gamma_{K^{*+} \rightarrow K^+ \gamma} &=& 50 \pm 5 \mbox{~keV} \;, \\
\Gamma_{K^{*0} \rightarrow K^0 \gamma} &=& 116 \pm 10 \mbox{~keV} \;.
\end{eqnarray}
In principle, one can determine the value of the magnetic transition
moment on the basis of the proportionality $\kappa_{K^* K}^2 \sim
\Gamma_{K^* \rightarrow K \gamma}$. Within the context of isobar
models, however, the coupling constants are frequently considered as
``effective couplings'' wherein, for example, part of final-state
interaction effects are absorbed.  It is a common procedure to use
only the ratios of the measured decay widths to connect isospin
related coupling constants. This leads to the following expression:
\begin{equation}
\frac{ \kappa^2_{K^{*0} K^0}}{ \kappa^2_{K^{*+} K^+}} =
\frac{\Gamma_{K^{*0} \rightarrow K^0 \gamma}}{\Gamma_{K^{*+}
\rightarrow K^+ \gamma}} \;,
\end{equation}
or:
\begin{equation}
\kappa_{K^{*0} K^0} = - 1.52\  \kappa_{K^{*+} K^+} \;.
\label{eq:K*+K*0}
\end{equation} 
The relative sign in the last expression was allocated on the basis of
a CQM prediction \cite{Singer}.

The nucleon magnetic transition moments are related to the
photohelicity amplitudes through the interaction Lagrangians. From the
isospin structure of the $N^*$ helicity amplitudes, it is easily
proven that they are sensitive to the isospin of the final state.  To
determine the electromagnetic vertex coupling at a neutron target from
the knowledge of the electromagnetic coupling at a proton target,
those differences have to be taken into account. We adopt the same
procedure as for the vector-meson transition moments and use the
experimental amplitudes as a conversion coefficient. The expressions,
which directly follow from the interaction Lagrangians, read:
\begin{eqnarray}
\mbox{spin-$\frac{1}{2}$:} &\hspace{0.1cm}& \frac{\kappa_{{\sss
N^*}n}}{ \kappa_{{\sss N^*}p}} = \frac{A^n_{1/2}}{A^p_{1/2}} \;, \\
\mbox{spin-$\frac{3}{2}$:} & & \frac{\kappa_{{\sss
N^*}n}^{\left( 1 \right)}}{ \kappa_{{\sss N^*} p}^{\left( 1 \right)}} =
\frac{ \sqrt{3} A^n_{1/2} \pm A^n_{3/2}}{\sqrt{3} A^p_{1/2} \pm
A^p_{3/2}} \;, \\
& &  \frac{\kappa_{{\sss N^*} n}^{\left( 2 \right)}}{ \kappa_{{\sss
N^*}p }^{\left( 2 \right)}} = \frac{ \sqrt{3} A^n_{1/2} -
\frac{M_p}{M_{\sss N^*}} A^n_{3/2}}{\sqrt{3} A^p_{1/2} -
\frac{M_p}{M_{\sss N^*}} A^p_{3/2}} \;,
\end{eqnarray}
where $\pm$ refers to even/odd parity.  Note that some of these
helicity amplitudes are rather poorly known, especially those of the
neutron.  For the electromagnetic decay of the $\Delta^*$ resonances
the following simple relation holds:
\begin{equation}
\kappa_{{\sss \Delta^*}p} = \kappa_{{\sss \Delta^*}n} \;,
\end{equation}
regardless of the spin state of the $\Delta^*$ resonance. This is not
the case for the electromagnetic decay of $\Sigma^*$ resonances. In
principle, one can make use of the same procedure adopted for the
$K^*$ and $N^*$ transition moment and take the ratio of the helicity
amplitudes as a conversion coefficient. Due to the lack of knowledge
about the latter quantities, we have used ratio's of the $\Sigma$
ground state transition moments as conversion coefficients. This
produces the following relations:
\begin{equation}
\kappa_{\Sigma^* \Sigma^0} = \frac{\mu_{\Sigma^0}}{\mu_{\Sigma^+}}
\kappa_{\Sigma^* \Sigma^+} = \frac{\mu_{\Sigma^0}}{\mu_{\Sigma^-}}
\kappa_{\Sigma^* \Sigma^- } \;, 
\end{equation}
in which we have used $\mu_{\Sigma^+}$ = 2.458, $\mu_{\Sigma^-}$ = -1.160
\cite{PDG} and $\mu_{\Sigma^0}$ = 0.79 \cite{Perkins}.

\begin{acknowledgments}
This work was supported by the Fund for Scientific Research - Flanders
under contract number 4.0061.99 and the Research Council of Ghent
University. 
\end{acknowledgments}

%\bibliography{../../refs/references}

\end{document}